\documentclass[a4paper,11pt]{article}
\usepackage{pos}
\usepackage{graphicx}

\title{Topological susceptibility, scale setting and 
universality from $Sp(N_c)$ gauge theories}
\ShortTitle{Topological susceptibility, scale setting and 
universality from $Sp(N_c)$ gauge theories}

\author*[a]{D.~Vadacchino}
\author[b]{E.~Bennett}
\author[c,d]{C.-J.~David~Lin}
\author[e]{D.~K.~Hong}
\author[f,e]{J.-W.~Lee}
\author[b,g]{B.~Lucini}
\author[h]{M.~Piai}

\affiliation[a]{Centre for Mathematical Sciences, University of Plymouth, Plymouth, PL4 8AA, United Kingdom}
\affiliation[b]{Swansea Academy of Advanced Computing, Swansea University (Bay Campus),\\Fabian Way, SA1 8EN Swansea, Wales, United Kingdom}
\affiliation[c]{Institute of Physics, National Yang Ming Chiao Tung University, 1001 Ta-Hsueh Road, Hsinchu 30010, Taiwan}
\affiliation[d]{Center for High Energy Physics, Chung-Yuan Christian University, Chung-Li 32023, Taiwan}
\affiliation[e]{Department of Physics, Pusan National University, Busan 46241, Korea}
\affiliation[f]{Institute for Extreme Physics, Pusan National University, Busan 46241, Korea}
\affiliation[g]{Department of Mathematics, Faculty of Science and Engineering, Swansea University (Bay Campus), Fabian Way, SA1 8EN Swansea, Wales, United Kingdom}
\affiliation[h]{Department of Physics, Faculty of Science and Engineering, 
Swansea University (Park Campus),\\Singleton Park, SA2 8PP Swansea, Wales, United Kingdom}

\emailAdd{davide.vadacchino@plymouth.ac.uk}
\emailAdd{e.j.bennett@swansea.ac.uk}
\emailAdd{dlin@nycu.edu.tw}
\emailAdd{dkhong@pusan.ac.kr}
\emailAdd{jwlee823@pusan.ac.kr}
\emailAdd{b.lucini@swansea.ac.uk}
\emailAdd{m.piai@swansea.ac.uk}

\abstract{
In this contribution, we report on our study of the properties
of the Wilson flow and on the calculation of the topological 
susceptibility of $Sp(N_c)$ gauge theories for $N_c=2,\,4,\,6,\,8$.
The Wilson flow is shown to scale according to
the quadratic Casimir operator of the gauge group,
as was already observed for $SU(N_c)$, and the 
commonly used scales $t_0$ and $w_0$ are obtained
for a large interval of the inverse coupling for 
each probed value of $N_c$.
The continuum limit of the topological susceptibility is computed
and we conjecture that it scales with the dimension of the group. 
The lattice measurements performed in the $SU(N_c)$ Yang-Mills theories
by several independent collaborations allow us to test this 
conjecture and to obtain a universal large-$N_c$ limit of the 
rescaled topological susceptibility.
}

\FullConference{%
The 39th International Symposium on Lattice Field Theory,\\
8th-13th August, 2022,\\
Rheinische Friedrich-Wilhelms-Universität Bonn, Bonn, Germany
}

\begin{document}
\maketitle

\section{Introduction}

Theories based on $Sp(N_c)$ gauge symmetry have been
under scrutiny in the past few years from different points of view,
prominently for their role in realizing UV-complete
composite Higgs models (CHMs) in which to implement top partial 
compositeness~\cite{Bennett:2019jzz,Bennett:2022yfa}.
In addition, they provide an alternative routes, besides the families
of groups $SU(N_c)$ and $SO(N_c)$, in the study of the universal features of the 
large-$N_c$ limit of gauge theories~\cite{Bennett:2020qtj,Bennett:2020hqd}. 
Several contributions relevant to this programme have been presented at 
this conference~\cite{Lee:2022elf,Zierler:2022qfq}, and a review of recent
results can be found in Ref.~\cite{Lucini:2021xke}.

This contribution reports on a lattice study of the topological features of the
vacuum of $Sp(N_c)$ pure gauge theories for $N_c=2$, $4$, $6$, and $8$. 
The scale of the lattice theory is set using the Wilson Flow (WF) and the
continuum and large-$N_c$ limits of topological susceptibility are obtained. 
In QCD, the latter 
enters crucially in the solution to the $U(1)_A$ problem~\cite{Witten:1979vv,Veneziano:1979ec}.
Moreover, it appears as the coefficient of the $O(\theta^2)$ term in the expansion of the 
the free energy in powers of a complex-valued $\theta$. Thus, it encodes properties of gauge
theories that might play a role in understanding the physics of the strong-CP 
problem (and the properties of the axion) and in the determination of the electric dipole 
moment of hadrons. Several studies have been devoted to this quantity 
for $SU(N_c)$ gauge theories~\cite{Lucini:2001ej,DelDebbio:2002xa,Bonati:2016tvi,Bonanno:2020hht,Athenodorou:2021qvs}. The comparison between these results and
the ones obtained for $Sp(N_c)$ suggest to propose a conjecture on the universal
properties of the topological susceptibility at large-$N_c$.
For additional details, see Refs.~\cite{Bennett:2022ftz,Bennett:2022gdz,bennett_ed_2022_6678411, bennett_ed_2022_6685967}.



\section{Lattice setup}\label{sec:lattice}

The lattice theory is defined on a four dimensional euclidean 
hypercubic lattice, with $L/a$ sites in each direction. Sites are labelled by 
$x$ and links by $(x,\mu)$.
The elementary degrees of freedom, defined on links, are $Sp(N_c)$-valued and 
denoted by $U_\mu(x)$. The action is defined as
\begin{equation}
	S_\mathrm{W}[U] = \beta\sum_{x,\mu>\nu} \left( 1 - 
\frac{1}{N_c} \mathrm{Re}\mathrm{Tr}~
      U_{\mu\nu}(x)
\right)~,\qquad\text{with}\quad\beta=\frac{2N_c}{g_0^2}~,
\end{equation}
where $g_0$ is the bare coupling and
$U_{\mu\nu}(x)=U_\mu(x) U_\nu(x+\hat{\mu}) U_\mu^\dag(x+\hat{\nu})U_\nu^\dag(x)$ is the 
so-called \emph{plaquette variable}. Expectation values of gauge invariant operators 
can be defined as finite integrals
in the measure $\mathcal{D}U_\mu=\prod_{x,\mu} dU_\mu(x)$, 
\begin{equation}
	\langle O \rangle = \frac{1}{Z} \int \mathcal{D}U_\mu O(U_\mu) e^{-S_\mathrm{W}[U]}
\end{equation}
where $Z=\int\mathcal{D}U_\mu  e^{-S_\mathrm{W}[U]}$, and computed numerically as \emph{ensemble averages}. 
The ensembles are sets of configurations of the gauge field 
generated with Monte Carlo sampling and stored for later analysis. 
Successive configurations are separated by a number $N_\mathrm{sw}$ of 
\emph{full lattice sweeps}, i.e. successive updates of all the 
links on the lattices with the traditional $1-4$ combination of 
heat bath (HB) and over-relaxation (OR) updates.
Ensembles of $\sim 4000$ configurations were obtained for various values of 
the input parameters $\beta$, $N_c$, and $L$. The specific values of the latter 
were chosen to reduce finite size effects. Details can be found
in Ref.~\cite{Bennett:2022ftz}. 

The difficulties in calculating the topological charge $Q$ and the topological
susceptibility $\chi$ from their lattice discretizations are well 
known~\cite{Panagopoulos:2011rb}. 
As a consequence of the discretization of the theory, the lattice topological charge 
$Q_L$ is not integer valued. Moreover, the lattice topological 
suspceptibility $\chi_L$ must be additively renormalized and is dominated by UV fluctuations
as $a\to 0$. In order to overcome these difficulties, and to set the scale of the lattice
at each value of $\beta$, the properties of the WF were 
employed~\cite{Luscher:2010iy,Luscher:2013vga}.
The flowed field $V_\mu(x,t)$ is defined by
\begin{equation}\label{eq:WFeq}
	\frac{\partial V_\mu(t,\,x)}{\partial t} = 
	-g_0^2 \left\{ \partial_{x,\mu} S_\mathrm{W}[V_\mu]\right\} V_\mu(t,\,x),\qquad
	V_\mu(0,\,x) = U_\mu(x)~,
\end{equation}
where $t$ is known as flow time. At large $t$, the WF drives the theory
to its classical limit and, at leading order in the coupling, 
$V_\mu(x,t)$ can be shown to be a gaussian smoothening of $U_\mu(x)$ with 
mean radius $1/\sqrt{8t}$. As a consequence of the suppression of UV fluctuations,
the lattice topological charge takes on quasi-integer values and the lattice 
topological susceptibility has an improved behaviour towards the continuum. 

Several definitions of the lattice topological charge density
are possible, that differ from one another by terms of order $O(a^4)$
as $a\to 0$.  
In this study, the clover-leaf expression was used
\begin{equation}\label{eq:clover}
q_L(x,\,t) \equiv \frac{1}{32\pi^2} 
\epsilon^{\mu\nu\rho\sigma}
\mathrm{Tr} 
~{\cal C}_{\mu\nu}(x,\,t) 
{\cal C}_{\rho\sigma}(x,\,t)\,,
\end{equation}
which is computed on $V_\mu(x,\,t)$. We refer to Ref.~\cite{Bennett:2022ftz} for the 
full expression of ${\cal C}_{\mu\nu}(x,\,t)$
as a function of $V_\mu(x,\,t)$.
The lattice topological charge can then be obtained as
$Q_L(t) = \sum_{x} q_L(x,\,t)$ and the effects of discretization can be further reduced by
employing the $\alpha$-rounding procedure~\cite{Bonati:2016tvi}, that yelds 
integer values for $Q_L$. 

Sampling different topological sectors at small lattice spacing with 
a local update 
algorithm is a notoriously difficult task, especially at larger values of $N_c$.
As $a\to 0$, the integrated autocorrelation time $\tau_Q$ of the $Q_L$ is known
to diverge exponentially~\cite{DelDebbio:2002xa}, with an exponent that increases with $N_c$.
Several remedies have been designed to restore ergodicity~\cite{Hasenbusch:2017unr,Bonanno:2022yjr,Eichhorn:2022wxn,Cossu:2021bgn} 
as $a\to 0$ at large-$N_c$. In this study, we tuned the value of 
$N_\mathrm{sw}$ so that, for each ensemble, $\tau_Q\sim 1$. 
That the distribution of the sampled topolofical charges was indeed
gaussian, as expected, was verified for each ensemble separately by 
performing a gaussian fit to the 
frequency histogram of $Q_L$. Finally, the topological susceptibility was evaluated as
\begin{equation}\label{eq:top_susc}
	\chi_L(t)a^{-4} = \frac{\langle Q_L^2(t) \rangle}{L^4}~.
\end{equation}
for each $N_c$, $\beta$ and $L$.

The scale of the lattice can be set by defining the renormalized coupling 
\begin{equation}
\label{eq:flowE}
    \alpha(\mu) \equiv k_\alpha t^2 \langle E(t) \rangle
	\equiv  k_\alpha \mathcal{E}(t)~,\quad 
	E(t)=\sum_x\tfrac{1}{2} \mathrm{Tr} V_{\mu\nu}(x,\,t) V_{\mu\nu}(x,\,t)
\end{equation}
at scale $\mu=1/\sqrt{8t}$, where $k_\alpha$ is a constant that can be calculated 
perturbatively, and $V_{\mu\nu}(x,\,t)$ is the plaquette variable calculated from $V_\mu(x,\,t)$. 
As an alternative way of setting the scale~\cite{Borsanyi:2012zs}, the quantity
\begin{equation}
\mathcal{W}(t) = t\frac{d}{dt}\left\{ \mathcal{E}(t) \right\}
\end{equation}
can be used instead
of $\mathcal{E}(t)$. As $\mathcal{W}(t)$ is expected to be affected by milder discretization effects than $\mathcal{E}(t)$, their comparison also provides an approximate 
way to quantify the magnitude of discretization
effects. The scales $t_0$ and $w_0$ can then be defined implicitly by $\mathcal{E}(t_0) = \mathcal{E}_0$ and
$\mathcal{W}(t_0) = \mathcal{W}_0$, where ${\cal E}_0$ and ${\cal W}_0$ can be chosen arbitrarily. A way to perform the 
large-$N_c$ limit at fixed 't~Hooft coupling, $\lambda=4\pi N_c \alpha$, 
can be understood from the perturbative expression
\begin{equation}\label{eq:scalingGF}
	\mathcal{E}(t) = \frac{3\lambda}{64\pi^2} C_2(F)~,
\end{equation}
where $C_2(F)$ is the quadratic Casimir operator of the gauge group. For the fundamental
representation of $Sp(N_c)$
one has $C_2(F)=(N_c+1)/4$. The above scaling law was tested in the context of $SU(N_c)$ gauge theories and 
good agreement was found with data, see Ref.~\cite{Ce:2016awn}. A constant 't~Hooft coupling is obtained at different
$N_c$ by choosing the reference values 
$\mathcal{E}_0$ and $\mathcal{W}_0$ according to,
\begin{equation}\label{eq:scaling_reference}
	\mathcal{E}_0 = c_e C_2(F),\qquad
	\mathcal{W}_0 = c_w C_2(F)~.
\end{equation}
where $c_e$ and $c_w$ are constants. While these tests are based on
perturbative arguments, we adopt the scaling in Eq.~(\ref{eq:scaling_reference})
throughout, to compare different theories.

\section{Numerical results}\label{sec:results}

\begin{figure}[t]
\centering
\includegraphics[width=0.7\textwidth]{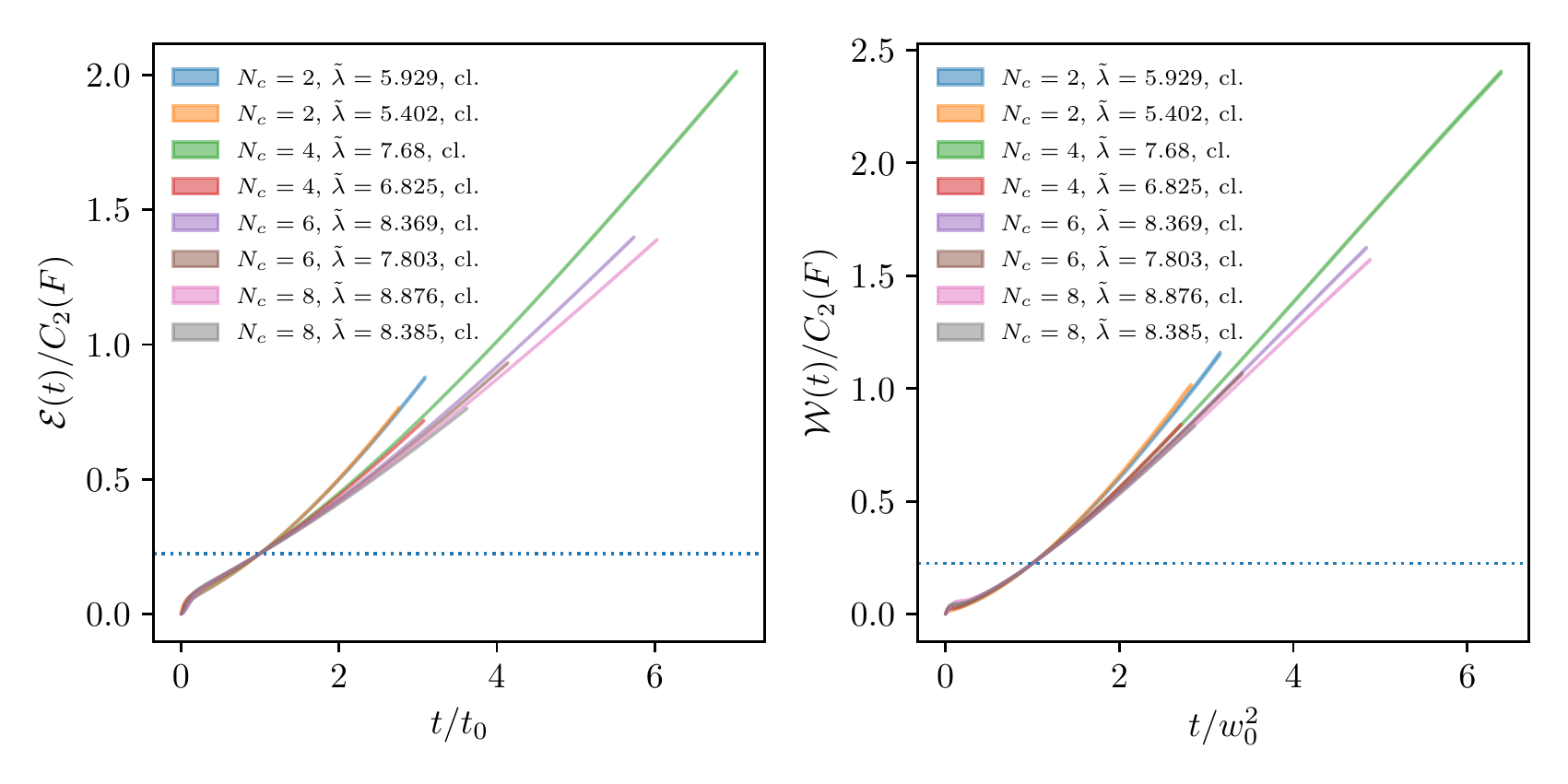}
\caption{
The quantities $\mathcal{E}(t)/C_2(F)$ (left panel) and
$\mathcal{W}(t)/C_2(F)$ (right), computed with the clover-leaf 
discretisation of $E(t)$ on the available ensembles corresponding to
the finest and coarsest available lattices, for each $N_c$, with $C_2(F)=(N_c+1)/4$, displayed as a function
of the rescaled flow times $t/t_0$ and $t/w_0^2$. The value of $\tilde\lambda$ 
is also reported in each case. The figure adopts the choice  $c_e=c_w=0.225$
(horizontal dashed line).
\label{fig:scaled_flows}}
\end{figure}

Each configuration in an ensemble was used as an initial condition
for the numerical integration of the WF equations, Eq.~(\ref{eq:WFeq}).
The quantity $\mathcal{E}(t)=t^2 E(t)$ was calculated on the entire interval
of $t$, using both the the plaquette (pl.) and the clover (cl.) 
expression for $V_{\mu\nu}$. 
The discretisation effects could be estimated and were only found to be 
non-negligible for a small neighbourhood of $t=0$. The value of $t_0$ and 
$w_0$ were determined at each $N_c$ from the reference 
values $\mathcal{E}_0$ and $\mathcal{W}_0$, obtained from Eq.~(\ref{eq:scaling_reference}), by setting $c_e=c_w=0.225$. Alternative scales were obtained by the choice $c_e=c_w=0.5$ 
and used for comparison. 
The behaviours of $\mathcal{E}(t)$ and $\mathcal{W}(t)$, normalized to $C_2(F)$,  are displayed
in Fig.~\ref{fig:scaled_flows}. They are expressed as functions of $t/t_0$ and $t/w_0^2$, and are
shown to overlap with each other over a wide range of values of $t$.
The overlap is more evident for ensembles that have a similar value 
of the convenently defined coupling 
$\tilde{\lambda}=\lambda d_G/\langle P \rangle$, where $\langle P \rangle$ is
the expectation value of the average action per plaquette; 
hence, the scaling law Eq.~(\ref{eq:scalingGF}) seems to capture some
scaling property that holds also non-perturbatively.

\begin{figure*}[t]
\centering
\includegraphics[width=.4\textwidth]{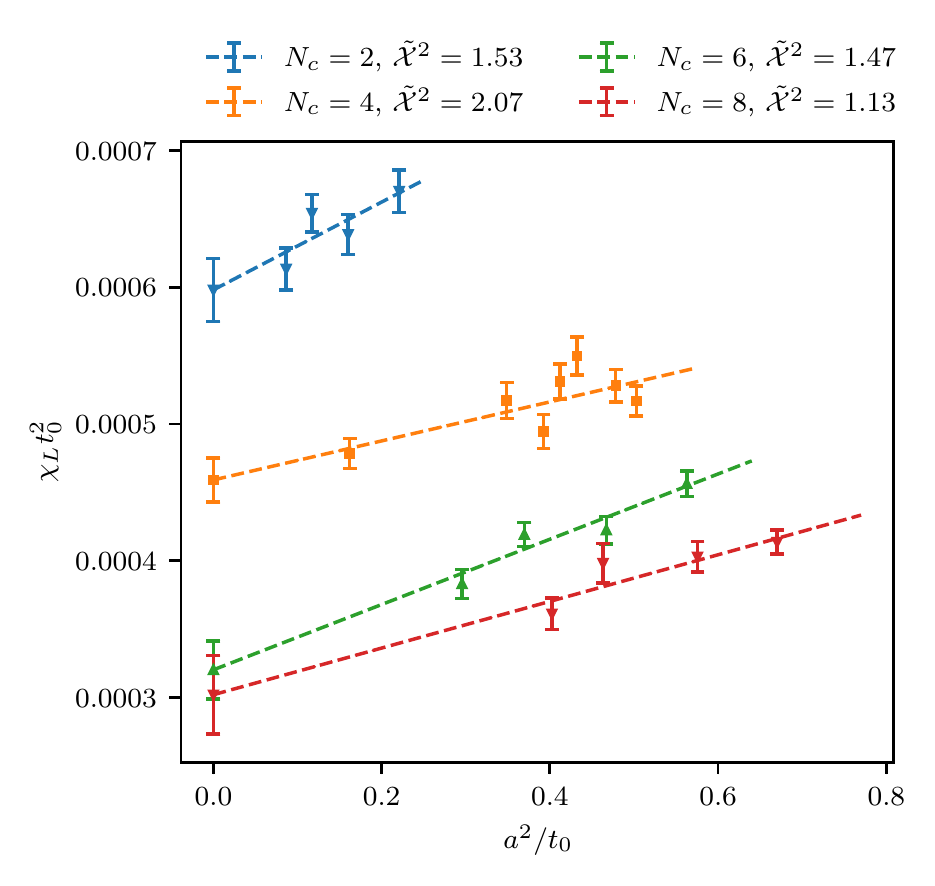}
\includegraphics[width=.4\textwidth]{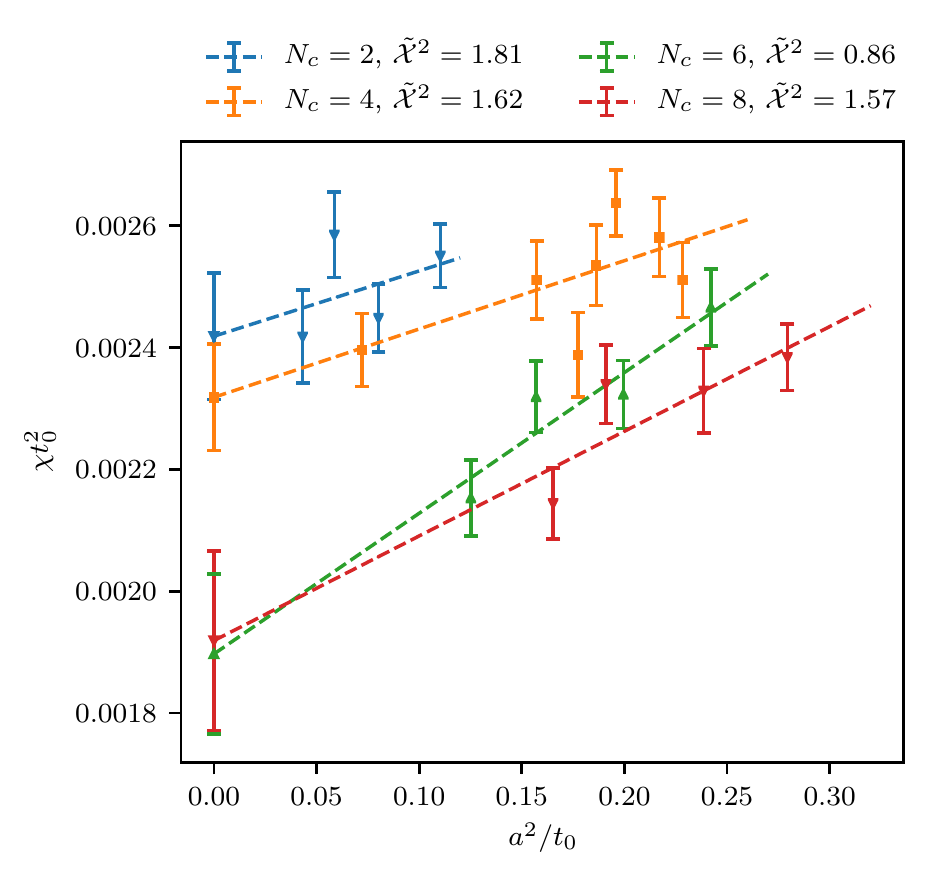}
\includegraphics[width=.4\textwidth]{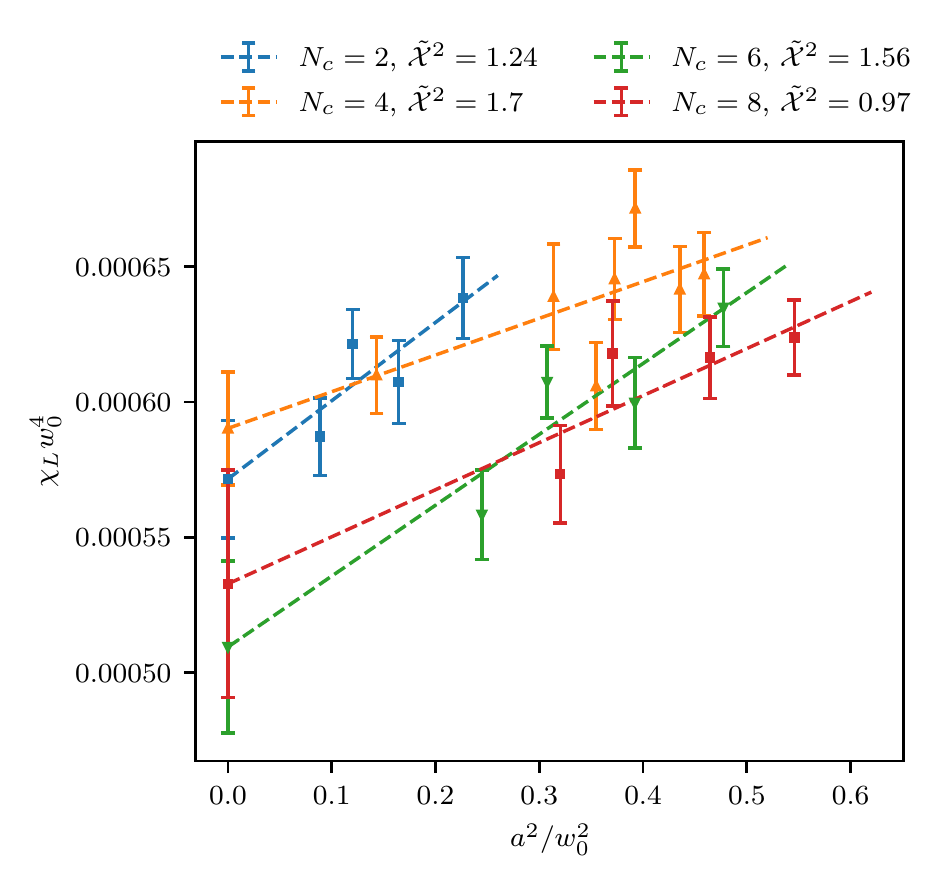}
\includegraphics[width=.4\textwidth]{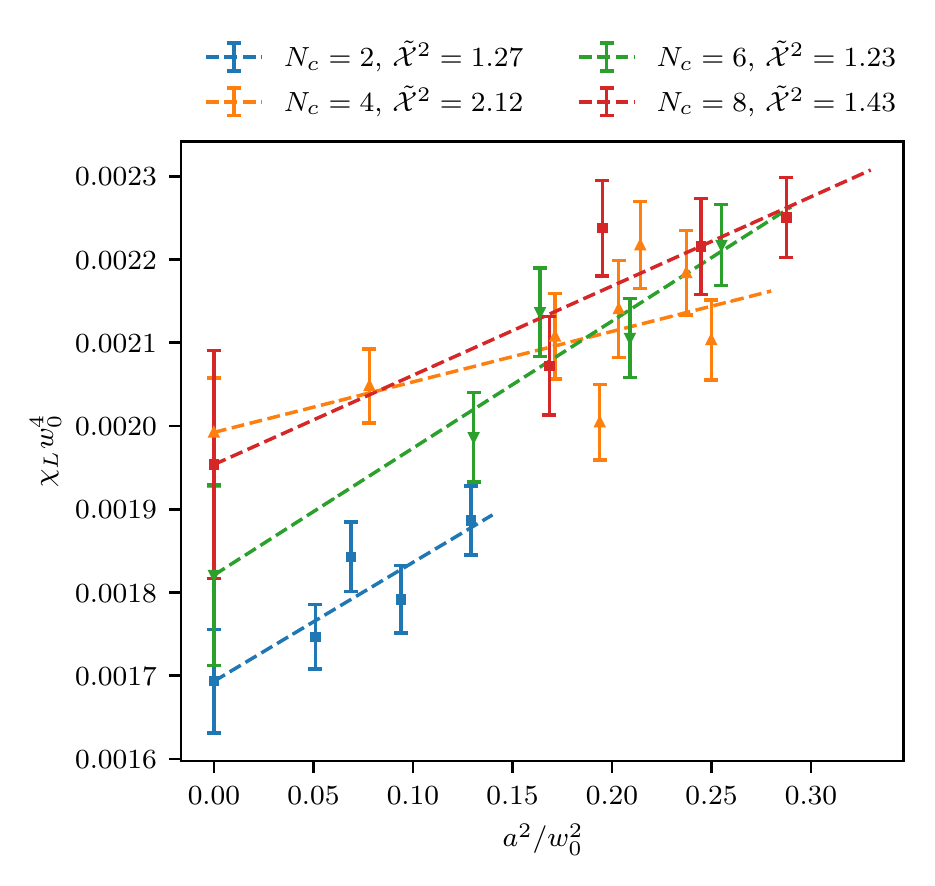}
\caption{Topological susceptibility per unit volume $\chi_L t_0^2$ as a
function of $a^2/t_0$ (top panels) and $\chi_L w_0^4$ as a
function of $a^2/w_0^2$ (bottom), in $Sp(N_c)$ Yang-Mills theories
with $N_c=2,\,4,\,6,\,8$. We adopt reference values  $c_e=c_w=0.225$ (left panels) and
$c_e=c_w=0.5$ (right).
Our continuum extrapolations are represented as dashed lines. The results are reported
in Table~\ref{tab:best_fits_contlim}.
\label{fig:best_fit_contlim}}
\end{figure*}

The $\alpha$-rounded lattice topological charge was compouted at $t=t_0$, 
where $t_0$ is obtained from Eq.~(\ref{eq:scaling_reference}) with both the choices $c_e=c_w=0.225$
and $c_e=c_w=0.5$, for comparison purposes. 
The values of the 
lattice topological susceptibility obtained from Eq.~(\ref{eq:top_susc}) are displayed 
in Fig.~\ref{fig:best_fit_contlim}, as a function of $a^2/t_0$ and $a^2/w^2_0$. The extrapolations to $a=0$ are performed according to a best fit of
\begin{equation}\label{eq:contlim}
\chi_L(a) t_0^2 = \chi_L(a=0) t_0^2 + c_1 \frac{a^2}{t_0}\,,
\end{equation}
to the data, using $c_1$ and $\chi_L(a=0)$ as fit parameters and are visible
as dashed lines. A similar analysis was carried out in units of $w_0$. 
The best-fit values are reported in Table~\ref{tab:best_fits_contlim}.

\begin{figure}[t]
\centering
\includegraphics[width=.4\textwidth]{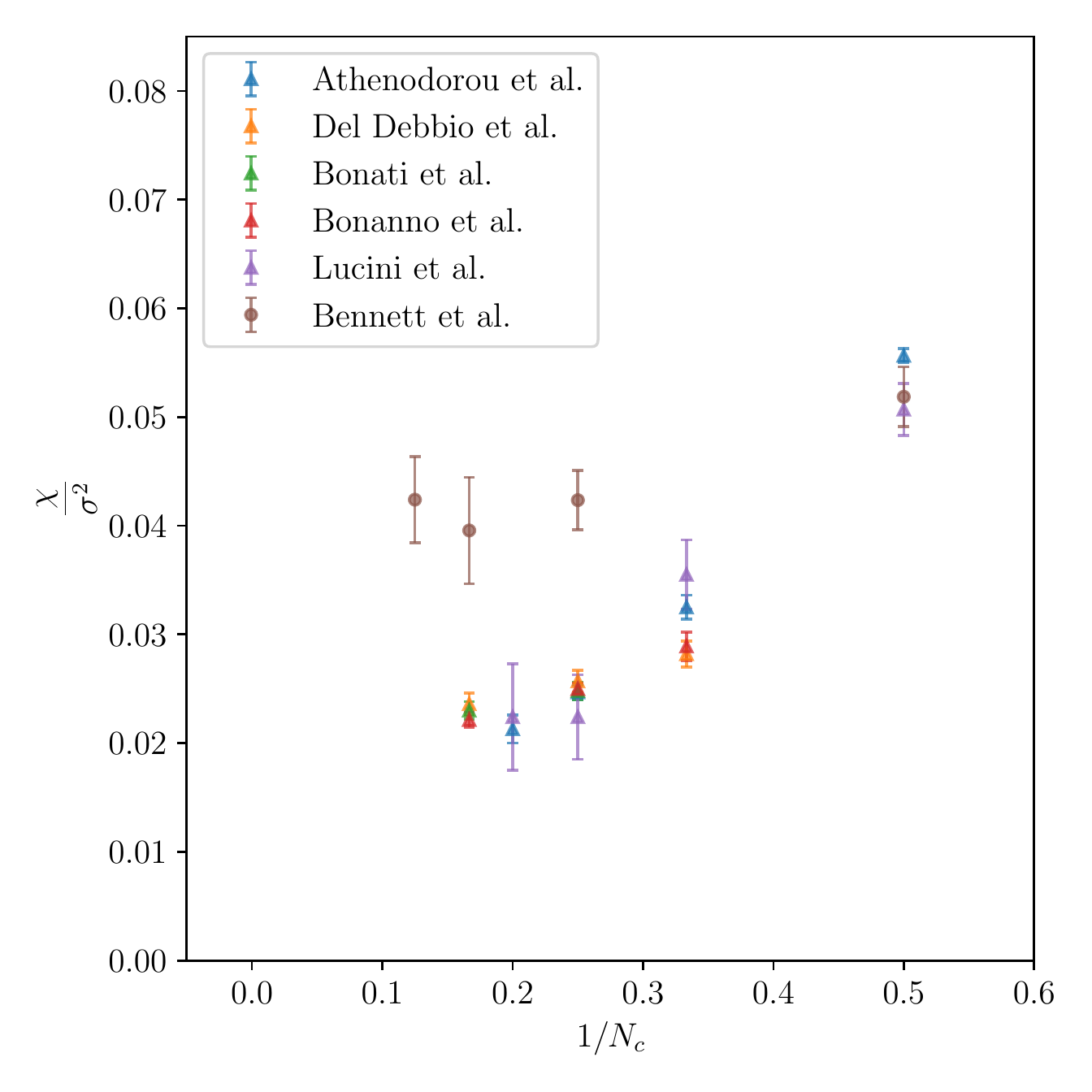}
\includegraphics[width=.4\textwidth]{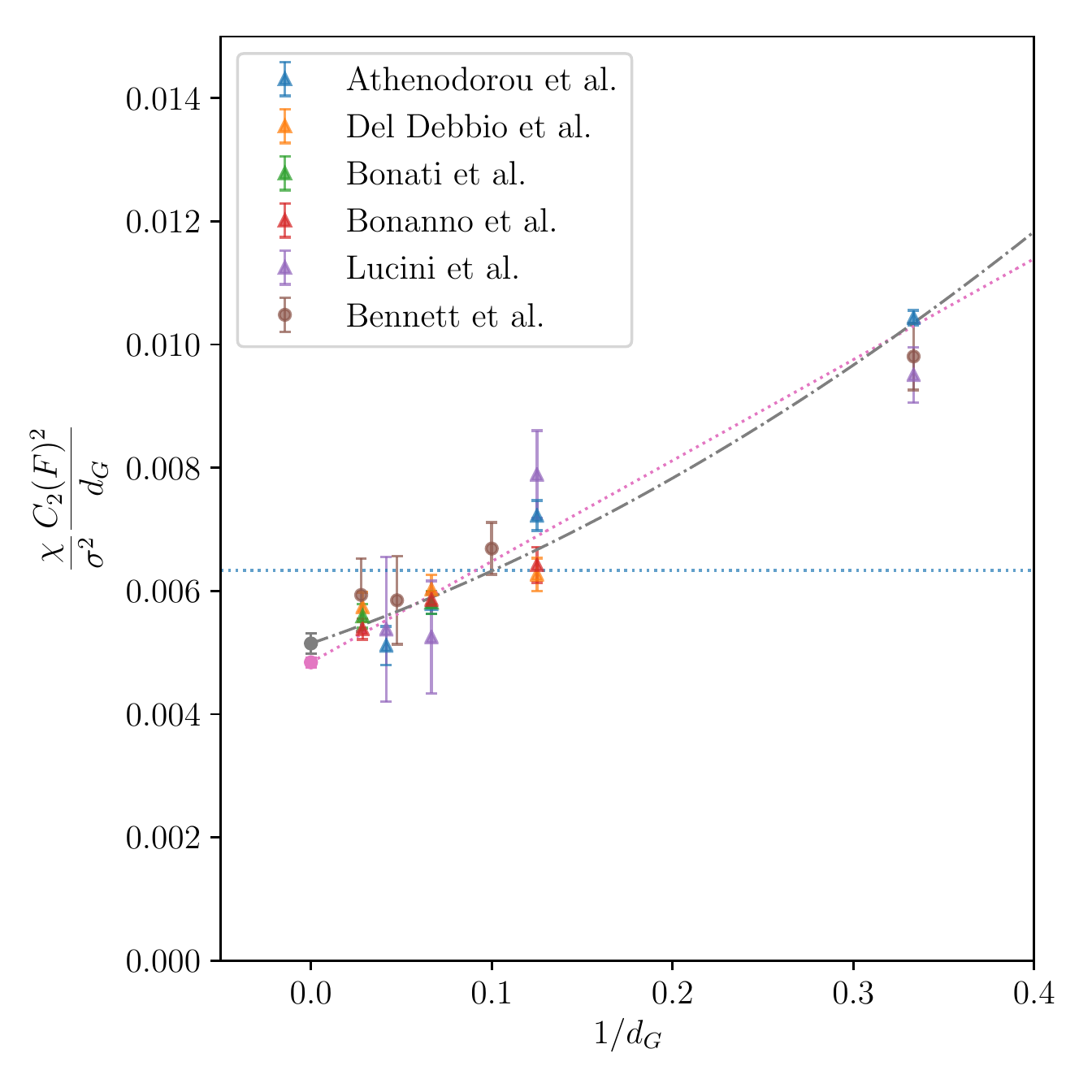}
\caption{Topological susceptibility $\chi$,
in units of the string tension $\sigma$, in the continuum limit, 
for various groups $SU(N_c)$ and $Sp(N_c)$, as a function of $1/N_c$(left panel)
and of $1/d_G$(right). 
The measurements reported here are labelled 
by the collaboration that published them. Figures are taken from Ref.~\cite{Bennett:2022gdz}
\label{fig:scaledchi}}
\end{figure}

Both the topological susceptibility and the string tension are assumed to tend to a 
finite limit as $N_c\to\infty$. As explained in Ref.~\cite{Bennett:2022gdz}, the 
gauge-group dependence 
of $\chi$ is inherited from the free energy of the theory. In particular, each of the $d_G$ gauge 
fields is expected to contribute equally to the latter. Moreover, assuming that the string tension 
$\sigma$ scales with the quadratic Casimir operator of the fundamental representation, 
we expect the following ratio to capture universal features:
\begin{equation}
\eta_{\chi}\equiv\frac{\chi C_2(F)^2}{\sigma^2 d_G} \,
=\frac{\chi}{\sigma^2}\cdot
\begin{cases}
\frac{N_c^2-1}{4N_c^2} & \text{ for } {SU}(N_c) \\
\frac{N_c+1}{8N_c} & \text{ for } {Sp}(N_c)
\end{cases}\,.
\end{equation}
In the limit $N_c\to\infty$, we expect that $\eta_X\to\eta_X(\infty)$, where
$\eta_X(\infty)$ is finite and universal. The latter was estimated
from the values of the topological susceptibility in 
$Sp(N_c)$ and $SU(N_c)$ at finite $N_c$, displayed in Fig.~\ref{fig:scaledchi} (left) as a function of $1/N_c$. The values obtained are displayed in 
Fig.~\ref{fig:scaledchi} (right) as a function of $1/d_G$. The result of a $2$-parameter
fit of $\eta_X(\infty)+b/d_G$ to the data, using $\eta_X(\infty)$ and $b$ as fitting parameters, 
is displayed as a dotted line. The estimate obtained for $\eta_X(\infty)$ is
\begin{equation}
	\eta_X(\infty)=(48.42\pm0.77\pm3.31)\times 10^{-4}~,
\end{equation}
where the first error is a statistical error produced by the fit, and the second is a systematic
error related to the choice of fitting function. The latter is estimated by the magnitude 
of the variation in the value of $\eta_X(\infty)$ if 
a term proportional to $1/d_G^2$ is added to the fitting function, 
and a $3$-parameters fit is performed.

\begin{table}
   \centering
\caption{The continuum limits of the topological susceptibility 
for $Sp(N_c)$ gauge theories with $N_c=2,\,4,\,6,\,8$, obtained from 
the best fit of Eq.~(\ref{eq:contlim}). The results are reported
in units of $t_0$ (top section of the table)
and in units of $w_0$ (bottom), obtained from the reference values
 $c_e=0.225=c_w$ (left section of the table)
or $c_e=0.5=c_w$ (right).
The value of the reduced chi-square for each extrapolation is labelled
as $\tilde{\mathcal{X}}^2$ and is reported in the last column.
The individual measurements and the extrapolations are displayed
in Fig.~\ref{fig:best_fit_contlim},
where each color corresponds to a different value of $N_c$.
\label{tab:best_fits_contlim}}
\begin{tabular}{|c|ccc|ccc|}
\hline
\hline
$N_c$ & $c_e$ & $\chi_L t_0^2(a=0)$  & $\tilde{\mathcal{X}}^2$ 
      & $c_e$ & $\chi_L t_0^2(a=0)$  & $\tilde{\mathcal{X}}^2$ \\
\hline
$2$ &$0.225$ & $0.000600(22)$ & $1.47$ &$ 0.5$  &$0.002353(93)$ & $1.29$ \\
$4$ &$0.225$ & $0.000452(17)$ & $2.18$ &$ 0.5$  &$0.002305(90)$ & $1.89$ \\
$6$ &$0.225$ & $0.000315(23)$ & $1.08$ &$ 0.5$  &$0.00185(11)$ & $1.02$ \\
$8$ &$0.225$ & $0.000303(23)$ & $2.32$ &$ 0.5$  &$0.00194(15)$ & $1.39$ \\
\hline
\hline
$N_c$ & $c_w$ & $\chi_L w_0^4(a=0)$  & $\tilde{\mathcal{X}}^2$ 
      & $c_w$ & $\chi_L w_0^4(a=0)$  & $\tilde{\mathcal{X}}^2$ \\
\hline
$2$ & $0.225$& $0.000572(24)$ & $1.16$ & $0.5$& $0.001698(69)$ & $1.50$ \\
$4$ & $0.225$& $0.000584(22)$ & $1.63$ & $0.5$& $0.001991(69)$ & $2.14$ \\
$6$ & $0.225$& $0.000503(32)$ & $1.39$ & $0.5$& $0.00180(12)$ & $1.13$ \\
$8$ & $0.225$& $0.000535(39)$ & $1.41$ & $0.5$& $0.00189(13)$ & $1.80$ \\
\hline
\end{tabular}
\end{table}

\section{Conclusion}\label{sec:conclusion}

This study focuses on the scaling properties of the WF and on the topological 
susceptibility in $Sp(N_c)$ gauge theories. 
Ensembles 
were collected at $N_c=2$, $4$, $6$ and $8$ over a range of values of the bare 
parameter $\beta$ and lattice volume $L$, chosen to avoid finite size effects.
An integer-valued lattice topological charge was obtained from each WF smoothed
configuration by $\alpha$-rounding. For each value of $N_c$, the topological 
susceptibility in units of $t_0$ and $w_0$ was then extrapolated to the 
continuum limit. The scales $t_0$ and $w_0$ were obtained at approximately
constant 't~Hooft coupling. The topological susceptibility for $Sp(N_c)$ 
and for $SU(N_c)$ gauge theories was then compared
at $N_c=\infty$, and their values used to test a universality conjecture.
The latter was found to be in agreement with the available data.

\acknowledgments

The work of E.~B. has been funded by the Supercomputing Wales project, which is part-funded by the European Regional Development Fund (ERDF) via Welsh
Government and by the UKRI Science and Technologies Facilities Council (STFC) Research Software Engineering Fellowship EP/V052489/1.
The work of D.~K.~H. was supported by
Basic Science Research Program through the National
Research Foundation of Korea (NRF) funded by the
Ministry of Education (NRF-2017R1D1A1B06033701).
The work of J.~W.~L is supported by the National Research 
Foundation of Korea (NRF) grant funded by the Korea government (MSIT) (NRF-2018R1C1B3001379).
The work of C.~J.~D.~L. and of H.~H. is supported
by the Taiwanese MoST Grant No. 109-2112-M-009-006-
MY3. The work of B.~L. and M.~P. has been supported in part
by the STFC Consolidated Grants No.~ST/P00055X/1 and No.~ST/T000813/1. B.~L. and M.~P. received funding from
the European Research Council (ERC) under the European
Union’s Horizon 2020 research and innovation program
under Grant Agreement No. 813942. The work of B.~L. is
further supported in part by the Royal Society Wolfson
Research Merit Award No. WM170010 and by the
Leverhulme Trust Research Fellowship No. RF-2020-4619. The work of D.~V. is supported in part by the INFN HPCHTC project and in part by the Simons Foundation under the
program “Targeted Grants to Institutes” awarded to the
Hamilton Mathematics Institute.  Numerical
simulations have been performed on the Swansea SUNBIRD
cluster (part of the Supercomputing Wales project) and AccelerateAI A100 GPU system,
on the local HPC clusters in Pusan National
University (PNU) and in National Chiao-Tung University
(NCTU), and on the Cambridge Service for Data Driven
Discovery (CSD3). The Swansea SUNBIRD system and AccelerateAI are part funded
by the European Regional Development Fund (ERDF) via
Welsh Government. CSD3 is operated in part by the
University of Cambridge Research Computing on behalf
of the STFC DiRAC HPC Facility (www.dirac.ac.uk). The
DiRAC component of CSD3 was funded by BEIS capital
funding via STFC capital Grants No. ST/P002307/1 and
No. ST/R002452/1 and STFC operations Grant No. ST/
R00689X/1. DiRAC is part of the National e-Infrastructure.

\bibliographystyle{JHEP}
\bibliography{proceeding.bib}

\end{document}